\documentclass{elsart}
\usepackage{epsfig}

\begin{document}

\begin{frontmatter}
  
\title{\boldmath Resonances in $J/\psi \to \phi \pi ^+\pi ^-$ and
    $\phi K^+K^-$}

\date{\today}

\maketitle

\begin{center}
\begin{small}
\vspace{0.2cm}

M.~Ablikim$^{1}$,              J.~Z.~Bai$^{1}$,               Y.~Ban$^{11}$,
J.~G.~Bian$^{1}$,  D.~V.~Bugg$^{20}$, X.~Cai$^{1}$,           J.~F.~Chang$^{1}$,
H.~F.~Chen$^{17}$,             H.~S.~Chen$^{1}$,              H.~X.~Chen$^{1}$
% ...
\\(BES Collaboration)\\

\vspace{0.2cm}
\label{att}
$^{1}$ Institute of High Energy Physics, Beijing 100049, People's Republic of China\\
$^{2}$ China Center for Advanced Science and Technology(CCAST), 
Beijing 100080, People's Republic of China\\
$^{3}$ Guangxi Normal University, Guilin 541004, People's Republic of China\\
$^{4}$ Guangxi University, Nanning 530004, People's Republic of China\\
$^{5}$ Henan Normal University, Xinxiang 453002, People's Republic of China\\
$^{6}$ Huazhong Normal University, Wuhan 430079, People's Republic of China\\
$^{7}$ Hunan University, Changsha 410082, People's Republic of China\\
$^{8}$ Liaoning University, Shenyang 110036, People's Republic of China\\
$^{9}$ Nanjing Normal University, Nanjing 210097, People's Republic of China\\
$^{10}$ Nankai University, Tianjin 300071, People's Republic of China\\
$^{11}$ Peking University, Beijing 100871, People's Republic of China\\
$^{12}$ Shandong University, Jinan 250100, People's Republic of China\\
$^{13}$ Shanghai Jiaotong University, Shanghai 200030, People's Republic of China\\
$^{14}$ Sichuan University, Chengdu 610064, People's Republic of China\\
$^{15}$ Tsinghua University, Beijing 100084, People's Republic of China\\
$^{16}$ University of Hawaii, Honolulu, Hawaii 96822, USA\\
$^{17}$ University of Science and Technology of China, Hefei 230026, People's Republic of China\\
$^{18}$ Wuhan University, Wuhan 430072, People's Republic of China\\
$^{19}$ Zhejiang University, Hangzhou 310028, People's Republic of China\\
$^{20}$ Queen Mary, University of London, London E1 4NS, UK \\
\vspace{0.4cm}

$^{a}$ Current address: Iowa State University, Ames, Iowa 50011-3160, USA.\\
$^{b}$ Current address: Purdue University, West Lafayette, IN 47907, USA.\\
$^{c}$ Current address: Cornell University, Ithaca, New York 14853, USA.\\
$^{d}$ Current address: Laboratoire de l'Acc{\'e}l{\'e}ratear Lin{\'e}aire, 
F-91898 Orsay, France.\\
$^{e}$ Current address: University of Michigan, Ann Arbor, Michigan 48109, USA.\\
$^{f}$ Current address: DESY, D-22607, Hamburg, Germany.\\
\end{small}
\end{center}

\normalsize

\begin{abstract}
A partial wave analysis is presented of
$J/\psi \to \phi \pi ^+\pi ^-$ and $\phi K^+K^-$
from a sample of 58M $J/\psi$ events in the BES\,II detector.
The $f_0(980)$ is observed clearly in both sets of data, and
parameters of the Flatt\' e formula are determined accurately:
$M = 965 \pm 8$ (stat) $\pm 6$ (syst) MeV/c$^2$,
$g_1 = 165 \pm 10 \pm 15 $ MeV/c$^2$,
$g_2/g_1 = 4.21 \pm 0.25 \pm 0.21$.
The $\phi \pi \pi$ data also exhibit a strong $\pi \pi$ peak centred at
$M = 1335$ MeV/c$^2$.
It may be fitted with $f_2(1270)$ and a dominant $0^+$ signal
made from $f_0(1370)$ interfering with a smaller $f_0(1500)$ component.
There is evidence that the $f_0(1370)$ signal is
resonant, from interference with $f_2(1270)$.
There is also a state in $\pi \pi$
with $M = 1790 ^{+40}_{-30}$  MeV/c$^2$ and $\Gamma = 270
^{+60}_{-30}$ MeV/c$^2$; spin 0 is preferred over spin 2.
This state, $f_0(1790)$, is distinct from $f_0(1710)$.
The $\phi K\bar K$ data contain a strong
peak due to $f_2'(1525)$.
A shoulder on its upper side may be fitted
by interference between $f_0(1500)$ and $f_0(1710)$.

\vspace{5mm}
\noindent{\it PACS:} 13.25.Gv, 14.40.Gx, 13.40.Hq

\end{abstract}

\end{frontmatter}
\clearpage

The processes $J/\psi \to \phi \pi ^+\pi ^-$ and $\phi K^+K^-$ have
been studied previously in the Mark III [1] and DM2 [2] experiments.
Here we report BES\,II data on these channels with much larger
statistics from a sample of 58 million $e^+e^- \to J/\psi$
interactions.  The $f_0(980)$, $f_0(1370)$ and a state with mass at
1790 MeV/c$^2$ and with spin 0 preferred over spin 2, called the $f_0(1790)$
throughout this paper, are studied here.  A particular feature is that
$f_0(1790) \to \pi \pi$ is strong, but there is little or no
corresponding signal for decays to $K\bar K$. This behavior is
incompatible with $f_0(1710)$, which is known to decay dominantly to
$K\bar K$; this indicates the presence of two distinct states, $f_0(1710)$
and $f_0(1790)$.

A detailed description of the BESII detector is given in Ref.
\cite{bes}.  It has a cylindrical geometry around the beam axis.
Trajectories of charged particles are measured in the vertex chamber
(VC) and main drift chamber (MDC); these are surrounded by a
solenoidal magnet providing a field of 0.4T.  Photons are detected in
a Barrel Shower Counter (BSC) comprized of a sandwich array of lead
and gas chambers. Particle identification is accomplished
using time-of-flight (TOF) information from the TOF scintillator array
located immediately outside the MDC and the $dE/dx$ information from the
MDC.

Events must have four charged tracks with total charge zero.  These
tracks are required to lie well within the MDC acceptance with a polar
angle $\theta$ satisfying $|\cos \theta | < 0.80$ and to have their
point of closest approach to the beam within 2 cm of the beam
axis and within 20 cm of the centre of the interaction region along
the beam axis.  Further, events must satisfy a four-constraint (4C)
kinematic fit with $\chi ^2 < 40$.

Kaons, pions, and protons are identified by time-of-flight, $dE/dx$,
and also by kinematic fitting.  The $\sigma$ of the TOF measurement is
180 ps.  Kaons may be identified by TOF and $dE/dx$ up to a momentum
of 800 MeV/c. The 4C kinematic fit provides additional good separation
between $\phi \pi \pi$ and $\phi KK$; residual crosstalk between these
channels is negligible.

The $K^+K^-$ invariant mass distributions for $J/\psi \to K^+K^-\pi
^+\pi ^-$ and $J/\psi \to K^+K^-K^+K^-$ are shown in Figs. 1(a) and
(c); in the latter case, the $K^+K^-$ pair with invariant mass closest
to the $\phi$ is plotted.  The peaks of the $\phi$ lie at 1019.7
$\pm$ 0.2 and 1020.0 $\pm$ 0.2 MeV/c$^2$ in (a) and (c), in reasonable
agreement with the value of the Particle Data Group (PDG) [4].  In
both cases, there is a clear $\phi$ signal over a modest background of
events due to $K^+K^-\pi ^+\pi ^-$ or $K^+K^-K^+K^-$ without a $\phi$.
The curves in (a) and (c) show the background, assuming it follows a
phase space dependence on $M(K^+K^-)$.  The resulting background is
$(19.0\pm1.5)$\% in (a) and $(6.2\pm1.6)$\% in (c).  Events containing
a $\phi$ are selected by requiring at least one kaon identified by TOF
or $dE/dx$ and $|M_{K^+K^-} - M_{\phi }| < 15$ MeV/c$^2$.

Before discussing the main physics results, it is necessary to deal
with an important background arising in $J/\psi \to K^+K^-\pi ^+\pi
^-$.  Events for the study of this background channel are selected in
a sidebin having $M(K^+K^-) = 1.045$--1.09 GeV/c$^2$.  Fig. 2 shows Dalitz
plots and mass projections for these sidebin events; Dalitz plots for
$\phi \pi ^+\pi ^-$ and $\phi K^+K^-$ data are shown in Fig. 3.  For
the $K^+K^-\pi ^+\pi ^-$ sidebin, there is a strong peak in the $\phi
\pi$ mass distribution of Fig. 2(b) centred at 1500 MeV/c$^2$ with a
full-width of 200 MeV/c$^2$.  This $\phi \pi$ peak is of interest because of
an earlier report of a possible exotic state close to this mass with
quantum numbers $J^P = 1^-$ [5].  The reflection of this peak produces
a horizontal band at the bottom of Fig. 2(a); it projects to a broad
peak centred at 2450 MeV/c$^2$ in Fig. 2(b).  For $K^+K^-K^+K^-$ sidebin
events of Fig. 2(c), there is no corresponding peak at low mass in
$\phi K$, Fig. 2(d).

In order to investigate the nature of this peak, we select events in
the mass range 1400--1600 MeV/c$^2$ from Fig. 2(b).  Mass distributions of
$K^+\pi ^-$ and $K^-\pi ^+$ pairs are shown in Fig. 2(e) and
corresponding distributions for $K^+\pi ^+$ and $K^-\pi ^-$ in Fig.
2(f).  There is a strong $K^*(890)$ peak visible in Fig. 2(e) but
only a broad peak in Fig. 2(f).
%
%
It can be shown that the presence of $K^*(890)$ in the background,
combined with kinematic selection in a narrow range of $K^+K^-$
masses, can generate the peak position and width of the spurious peak
in $\phi \pi$.

A similar effect arises in selected $\phi \pi \pi$ events.  Fig. 3(b)
shows $M(\phi \pi)$ for events selected as $\phi
\pi^+\pi ^-$ by requiring $M(K^+K^-)$ within $\pm 15$ MeV/c$^2$ of the
$\phi$ mass.  There is again a $\phi \pi$ peak, centred now at 1460
MeV/c$^2$.
Again it can be shown that the peak is consistent
entirely with background. There is no
significant evidence for an exotic $\phi \pi$ state.  If it were
misinterpreted as a $\phi \pi$ state, fits show that it requires a
$\phi$ combined with an $L = 1$ pion coming from the $K^*_1(890)$,
hence quantum numbers $J^P = 1^-$, $0^-$, or $2^-$.

We have carried through a full
partial wave analysis in three alternative ways: (a) making a cut in
order to exclude events lying within $\pm 80$ MeV/c$^2$ of the central
mass of $K^*(890)$, which is slightly narrower than the selection of
Fig. 1; (b) including into the fit an incoherent background from
$K^*(890)K\pi $; and (c) making a background subtraction which allows for
the shift in mass and width between sidebin and data for the
background peak in $\phi \pi$ at 1500 MeV/c$^2$.  Results of these
three approaches agree within errors.  We regard the first method as
the most reliable, since it is independent of any modelling of the
background. Figs. (4) and (5) show the fit from this approach.  The
cut against $K^*(890)$ eliminates the $\phi \pi$ peak at 1460
MeV/c$^2$, as shown in Fig. 4(d).  It also eliminates backgrounds due
to channels $K^*(1430)K^*(890)$, observed in the final state
$K^+K^-\pi ^+\pi ^-$.  It reduces the background under the $\phi$ in
Fig. 1(e) to $(13.5\pm1.4)$\%; after the background subtraction, the
number of $\phi \pi ^+\pi ^-$ events falls to 4180.

The branching fractions for production of $\phi \pi \pi$ and $\phi K\bar
K$ are determined allowing for the efficiencies for detecting the two
channels and correcting for unobserved neutral states.  Results are:
$B(J/\psi \to \phi \pi \pi ) = (1.63 \pm 0.03 \pm 0.20 ) \times
10^{-3}$ and $B(J/\psi \to \phi K \bar K ) = (2.14 \pm 0.04 \pm 0.22
) \times 10^{-3}$.  The main contributions to the systematic errors
come from differences between data and Monte Carlo simulation for the
$\phi$ selection, $K^*(890)$ cut, and particle identification;
from uncertainties in the MDC wire resolution; and the total number of
$J/\psi$ events.

We turn now to the physics revealed by diagonal bands in the Dalitz plots of
Fig. 3 and mass projections of Figs. 4 and 5.
There is a strong $f_0(980) \to \pi ^+\pi ^-$ signal in Fig. 4(c)
and a low mass peak in Fig. 5(c) due to $f_0(980) \to K^+K^-$.
Secondly, the $\phi \pi ^+\pi ^-$ data exhibit in Fig. 4(c) the clearest
signal yet observed for $f_0(1370) \to \pi ^+\pi ^-$.
Several authors have previously expressed doubts concerning the existence of
$f_0(1370)$, but present data cannot be fitted adequately without it.
Both Mark III and DM2 groups observed a similar peak with lower
statistics.
There have been earlier reports of similar but less conspicuous peaks
in $\pi \pi \to K\bar K$ from experiments at ANL [6,7] and BNL [8].
A third feature in the  $\phi \pi ^+\pi ^-$ data in Fig. 4(c) is
a clear peak at around 1775 MeV/c$^2$.

The $\phi K^+K^-$ data of Fig. 5(c) contain a strong $f_2'(1525)$
peak.  However, it is asymmetric and may only be fitted by including
on its upper side $f_0(1710)$ interfering with other components.

We now describe the maximum likelihood fit to the data.
Amplitudes are fitted to relativistic tensor expressions which are
documented in Ref. \cite{pwa}.
The full angular dependence of decays of the $\phi$ and $\pi ^+\pi ^-$
or $K^+K^-$ resonances is fitted, including correlations between them.
The line-shape of the $\phi$ is not fitted, because the
$\phi$ is much narrower than the experimental resolution.
We include production of $J^P = 0^+$ resonances with orbital angular
momentum $\ell = 0$ and 2 in the production process
$J/\psi \to \phi f_0$.
For production of $f_2$, there is one amplitude with
$\ell = 0$ and three with $\ell = 2$, where $\ell$ and the spin of $f_2$
may combine to make overall spin $S = 0$, 1 or 2.
The one possible $\ell = 4$ amplitude makes a negligible contribution.
The acceptance, determined from a Monte Carlo simulation, is included
in the maximum likelihood fit.
All figures shown here are uncorrected for acceptance, which is
approximately uniform
across Dalitz plots except for the effect of the $K^*(890)$ cut.

The background subtraction is made by giving data positive
weight in log likelihood and sidebin events negative weight; the
sidebin events (suitably weighted by $K^+K^-$ phase space) then cancel
background in the data sample.

The $\phi \pi ^+\pi ^-$ and $\phi K^+K^-$ data are fitted
simultaneously, constraining resonance masses and widths to be
the same in both sets of data.
Table 1 shows branching fractions of each component,
as well as the
changes in log likelihood when each component is
dropped from the fit and remaining components are re-optimised.

We begin the discussion with $\phi K^+K^-$ data.  There is a
conspicuous peak due to $f_2'(1525)$. The shoulder on its upper side
is fitted mostly by $f_0(1710)$ interfering with $f_0(1500)$, but
there is also a possible small contribution from $f_0(1790)$
interfering with $f_0(1500)$. The overall contributions to $\phi
K^+K^-$ are shown by the upper histograms in Figs 5(c) and (d).

The $f_2(1270)$ signal reported below in $\phi \pi \pi$ data
allows a calculation of the $f_2(1270) \to K^+K^-$  signal
expected in $\phi K^+K^-$, using the
branching fraction ratio between $K\bar K$ and $\pi \pi$ of the PDG. Its
contribution is negligibly small.

We discuss next the fit to $f_0(980)$.
In $\phi \pi ^+\pi ^-$ data, it interferes with a broad
component well fitted by the $\sigma$ pole \cite{opipi}.
This component interferes constructively with the lower side of the
$f_0(980)$ in Fig. 4(c). Its magnitude is shown by the lower curve in
Fig. 4(e).

The $f_0(980)$ amplitude  has
been fitted to the Flatt\' e form:
\begin {equation} f=\frac {1}{M^2 - s - i(g_1\rho _{\pi \pi } +
g_2\rho _{K\bar K})}.
\end {equation}
Here $\rho$ is Lorentz invariant phase space, $2k/\sqrt {s}$, where $k$
refers to the $\pi$ or $K$ momentum in the rest frame of the
resonance.
The present data offer the opportunity to determine the
ratio $g_2/g_1$ accurately.
This is done by determining the number of
events due to $f_0(980) \to \pi \pi$ and $\to K^+K^-$ and comparing
with the prediction from the Flatt\' e formula, as follows.
After making the best
fit to the data, the fitted $f_0(980) \to \pi ^+\pi ^-$
signal is integrated over the mass range from 0.9 to 1.0 GeV/c$^2$.
The fitted $f_0(980) \to K^+K^-$ signal is integrated over the mass
range 1.0--1.2 GeV/c$^2$, so as to avoid sensitivity to the tail of the
$f_0(980)$ at high mass.
The latter integral is given by
\begin {equation} 0.5 \int ds |f(980)|^2\rho(K^+K^-)\epsilon(K^+K^-)
\end {equation}
and the former by
\begin {equation}
\frac {2}{3} \int ds |f(980)|^2\rho(\pi \pi )\epsilon(\pi ^+\pi ^-).
\end {equation}
Here $\epsilon(K^+K^-)$ and $\epsilon(\pi ^+\pi ^-)$ are detection
efficiencies.  The numerical factors at the beginning of each
expression take into account (a) there are equal
numbers of decays to $K^+K^-$ and $K^0\bar K^0$ and (b) two-thirds of
$\pi \pi $ decays are to $\pi ^+\pi ^-$ and one third to $\pi ^0 \pi
^0$.

By an iterative process which converges rapidly, the ratio $g_2/g_1$
is adjusted until the ratio of these two integrals reproduces the
fitted numbers of events for $\phi K^+K^-$ and $\phi \pi ^+\pi ^-$.
The result is $g_2/g_1 = 4.21 \pm 0.25$ (stat) $\pm 0.21$ (syst).
The systematic error arises from (i) varying the choice of
side bins and the magnitude of the background under the $\phi$ peak,
(ii) changes in the fit when small amplitudes such as
$f_0(1500)$ and $f_2(1270) \to K^+K^-$ and $\sigma\to K^+K^-$
are omitted from the fit,
(iii) changing the mass and width of other components within errors and
different choices of $\sigma$ parameterization from Ref. \cite{opipi}.
The result is a considerable
improvement on earlier determinations. The mass and $g_1$ are adjusted
to achieve the best overall fit to the peak in $\phi \pi ^+\pi ^-$
data. Values are $M = 965 \pm 8 \pm 6$ MeV/c$^2$,
$g_1 = 165 \pm 10 \pm 15$ MeV/c$^2$.

The ratio $g_2/g_1$ is only weakly correlated with $M$ and $g_1$.
However, $g_2 $ is rather strongly correlated with $M$.  This arises
because the term $ig_2\rho _{KK}(s)$ in the Breit-Wigner denominator,
eqn. (1), continues analytically below the $KK$ threshold to
$-g_2\sqrt {(4M^2_K/s) - 1}$.  It then contributes to the real part of
the Breit-Wigner amplitude and interacts with the term $(M^2 - s)$. We
find that the correlation is given by $dg_2/dM = -5.9$; the mass goes
down as $g_2 $ goes up.  Other correlations are weak: $dg_1/dM =
-0.75$ and $dr/dg_1 = -0.068$, where $r = g_2/g_1$.

We consider next the peak in $\phi \pi \pi$ centred at a mass of 1335
MeV/c$^2$.  An initial fit was made to $f_2(1270)$ and one $f_0$.  The
$f_0$ optimizes at $M=1410 \pm 50$ MeV/c$^2$, $\Gamma = 270 \pm 45$
MeV/c$^2$, where errors cover systematic variations when small
ingredients in the fit are changed. However, both $f_0(1500)$ and
$f_0(1370)$ can contribute.  Adding $f_0(1500)$, log likelihood
improves by 51: an 8.5 standard deviation improvement for four degrees
of freedom.  Also the fit to the $\pi \pi$ mass distribution improves
visibly.  Therefore three components are required in the 1335 MeV/c$^2$
peak: $f_2(1270)$, $f_0(1370)$ and $f_0(1500)$.  Removing $f_0(1370)$
makes log likelihood worse by 83, a 10.8 standard deviation effect.

Angular correlations between decays of $\phi $ and $f_2$ are very
sensitive to the presence of $f_2(1270)$, which is accurately
determined. It optimizes at $M = 1275 \pm 15$ MeV/c$^2$,
$\Gamma = 190 \pm 20$ MeV/c$^2$, values consistent with $f_2(1270)$.
The fact that its mass and width agree well with PDG values rules
out the possibility that the remainder of the signal in this mass
range is due to spin 2; otherwise the fit to $f_2(1270)$ would
be severely affected. Angular distributions for the remaining 
components are indeed consistent with isotropic decay angular 
distributions from spin 0.

The $f_0(1370)$ interferes with $f_0(1500)$ and $f_2(1270)$.  This
helps to make $f_0(1370)$ more conspicuous than in other data.
However, because of the interferences,
its mass and width are not accurately determined.
The mass of $f_0(1370)$ is $1350 \pm 50$ MeV/c, where the error is 
the quadratic sum of the systematic and statistical errors.

The width of $f_0(1370)$ is somewhat more stable. It is determined
essentially by the full width of the peak in $\phi \pi ^+ \pi ^-$ of
270 MeV/c$^2$; interferences with $f_2(1270)$ and $f_0(1500)$ affect this
number only by small amounts and the fitted width is $265 \pm 40$ MeV/c$^2$.
If both $f_0(1370)$ and $f_0(1500)$ are removed, log likelihood is
worse by 595.  Removing $f_0(1500)$ from the fit perturbs the mass
fitted to $f_0(1370)$ upwards to 1410 $\pm$ 50 MeV/c$^2$; this is
obviously due to the fact that $f_0(1370)$ is trying to simulate the
missing $f_0(1500)$ component.

The presence of a peak due to $f_0(1370)$ is strongly suggestive of a
resonance. In order to check for resonant phase variation, we have
tried replacing the amplitude by its modulus, without any phase
variation. In this case, log likelihood is worse by 39, nearly a 9
standard deviation effect for a change of one degree of freedom.  The
conclusion is that the $f_0(1370)$ peak is resonant.  It is not
possible to display the phase directly, since it is determined by
interferences between two $f_0(1370)$ and four $f_2(1270)$ amplitudes.

The magnitude of the signal due to $f_0(1370) \to K^+K^-$
in the fit gives a branching fraction ratio
\begin {equation}
\frac {B[f_0(1370) \to K\bar K]}{B[f_0(1370) \to \pi \pi]} = 0.08\pm 0.08.
\end {equation}
This value is somewhat lower than reported by the Particle Data
Group [4]. The reason is the conspicuous signal in $\pi \pi$ but
absence of any corresponding peak in $K^+ K^-$.

Next we consider the peak in $\pi ^+\pi ^-$ at 1775 MeV/c$^2$ in Fig.
4(c).  It fits well with $J^P = 0^+$ with $M = 1790 ^{+40}_{-30}$
MeV/c$^2$, $\Gamma = 270 ^{+60}_{-30}$ MeV/c$^2$.  The fitted mass is
in reasonable accord with the $f_0(1770)$ reported in Crystal Barrel
data on $\bar pp \to (\eta \eta )\pi ^0$~\cite{crystal}: $M = 1770 \pm
12$ MeV/c$^2$, $\Gamma = 220 \pm 40$ MeV/c$^2$.  Allowing for the
number of fitted parameters, $f_0(1790)$ is more than a $15\sigma$
signal.  It cannot arise from $f_0(1710)$, since the magnitude of
$f_0(1710) \to K^+K^-$ is small (see Table 1), and it is known that
the branching fraction ratio of $f_0(1710)$ between $\pi \pi$ and
$K\bar K$ is $<0.11$ at the $95\%$ confidence level \cite{okk};
accordingly, the $f_0(1710) \to \pi \pi$ signal in present data should
be negligibly small.

We now consider possible fits with an $f_2$ instead.
The decay angular distribution in this mass range is consistent with
isotropy. So there is no positive evidence for spin 2.
However, four spin 2 amplitudes are capable of simulating
a flat angular distribution.
In consequence, spin 2 gives a log likelihood which is worse than
spin 0 by only 4.5 after re-optimising its mass and width.
If $f_0(1710)$ is then added with PDG mass and
width, it improves log likelihood by a further 2.0; this
confirms the result from $\omega K^+K^-$ data that
$f_0(1710)$ has a negligible decay to $\pi \pi$.
Our experience elsewhere is that
using four helicity amplitudes instead of 2 adds considerable
flexibility to the fit.
The spin 2 amplitude with $\ell = 0$ has a distinctive term
$3\cos ^2 \alpha _\pi - 1$, where $\alpha _\pi$ is the
decay angle of the $\pi^+$ in the resonance rest frame, with respect
to the direction of the recoil $\phi$.
Simulation of spin 0 requires large $J = 2$, $\ell = 2$ and 4
amplitudes to produce compensating terms in $\sin ^2\alpha_{\pi}$
and hence a flat angular distribution.
Large contributions from $\ell = 2$ are
unlikely in view of the low momentum available to the resonance
and the consequent $\ell = 2$ centrifugal barrier.
If the $J = 2$ hypothesis is fitted only with $\ell = 0$,
log likelihood is worse by 95 than for spin 0.
We conclude that the state is most likely spin zero.

It is not possible to fit the shoulder in $\phi K^+K^-$ at 1650 MeV/c$^2$
accurately by interference between $f_0(1500)$ and $f_0(1790)$, using
the $f_0(1790)$ mass and width found in $\phi \pi \pi$ data.  Even if one accepts
the poor fit this gives, the branching fraction ratio $K\bar K$/$\pi \pi$
assuming only one $f_0$ resonance here is $0.55 \pm 0.10$.  This is a factor
14 lower than that reported in Ref. \cite{okk} for $f_0(1710)$.  For a
resonance, branching fractions must be independent of production
mechanism. The large discrepancy in branching fractions implies the
existence of two distinct states at 1710 and 1790 MeV/c$^2$, the former
decaying dominantly to $K\bar K$ and the latter dominantly to $\pi
\pi$.  The $f_0(1790)$ is a natural candidate for the radial
excitation of $f_0(1370)$.  There is earlier evidence for it decaying
to $4\pi$ in $J/\psi \to \gamma (4\pi)$ data [13,14], with mass and
width close to those observed here.  There, spin 0 was preferred
strongly over spin 2.

The shoulder in $\phi K^+K^-$ at 1650 MeV/c$^2$ is fitted
with interference between $f_0(1500)$ and $f_0(1710)$, which is known
to decay strongly to $K\bar K$. If both  $f_0(1710)$ and $f_0(1790)$
are included in the fit, there is only a small improvement from
$f_0(1790)$.

Masses, widths and branching fractions are given in Table I.
The errors arise mainly from
(i) varying the choice of side bins and the magnitude of the background
under the $\phi$ peak,
(ii) adding or removing small components
such as $f_0(1500)$, $f_2(1270) \to K^+K^-$, and $\sigma\to K^+K^-$ and
(iii) varying the mass and width of every component within errors and
using different $\sigma$ parameterizations reported in Ref. \cite{opipi}.
It also includes the uncertainty in the number of $J/\psi$ events
and the difference between two alternative choices of MDC wire
resolution simulation.

\begin {table}[htp]
\begin {center}
\begin {tabular}{|cccccc|}
\hline
Channel & Mass & Width & $B(J/\psi\to\phi X,$  & $B(J/\psi\to\phi X,$ & $\Delta S$\\
        & (MeV/c$^2$)& (MeV/c$^2$) &  $X\to\pi\pi)$ & $X\to K \bar{K})$ & \\
        &      &       & $(\times 10^{-4})$ & $(\times 10^{-4})$ & \\\hline
$f_0(980)$   & $965 \pm 10 $ & see text    & $5.4\pm0.9$ & $4.5\pm0.8$ & 1181\\
$f_0(1370)$  & $1350\pm 50$ & $265\pm40$   & $4.3\pm1.1$ & $0.3\pm0.3$ & 83  \\
$f_0(1500)$  & PDG          & PDG          & $1.7\pm0.8$ & $0.8\pm0.5$ & 51 \\
$f_0(1790)$  & $1790^{+40}_{-30}$ & $270^{+60}_{-30}$ & $6.2\pm1.4$ &  $1.6\pm 0.8$ & 488\\
$f_2(1270)$  & $1275\pm15$ & $190\pm20$    & $2.3\pm0.5$ &  $0.1\pm0.1$   & 241\\
$\sigma$     &              &              & $1.6\pm0.6$ &  $0.2\pm0.1$   & 120 \\
$f_2'(1525)$ & $1521 \pm 5$ & $77\pm15$    &  -   & $7.3\pm1.1$ & 440\\
$f_0(1710)$  & PDG & PDG                   &  -   & $2.0\pm0.7$ & 64\\
\hline
\end {tabular}
\caption {Parameters of fitted resonances and branching fractions for each channel;
improvements in $S=$ log likelihood when the channel is added.  PDG means
that the mass and width are fixed to the PDG value. For the
$f_0(980)$, see the parameterization in the text. The errors are the
statistical and systematic errors added in quadrature.}
\end {center}
\end {table}

Finally,
angular distributions for both production and decay have been examined
for each resonance peak. There are no significant discrepancies between
data and fit.
A fit is shown for the $f_0(1790)$ peak in Fig. 6 to the decay angle
$\alpha _\pi$ of the $\pi\pi$ pair, with respect to the recoil $\phi$;
the deep dip at $\cos \alpha _\pi =  \pm 0.75$ is due to the $K^*(890)$
cut. The remaining angular distribution fits well to spin 0.

It is remarkable that $\phi \pi \pi$ data contain large signals
due to several states which are predominantly non-strange:
$f_2(1270)$, $f_0(1370)$, $f_0(1500)$ and $f_0(1790)$;
direct production with the $\phi$ should favour $s\bar s$ states.
There is no agreed explanation.

In summary, the data reported here have three important features.
Firstly, the parameters of $f_0(980)$ are all well determined.
Secondly, there is the clearest signal to date of $f_0(1370) \to \pi
^+\pi ^-$; a resonant phase variation is required, from
interference with $f_2(1270)$.  Thirdly, there is a clear peak in $\pi
\pi$ at 1775 MeV/c$^2$, consistent with $f_0(1790)$; spin 2 is less likely
than spin 0.  If the $f_0(1790)$ resonance is used to fit the shoulder
at 1650 MeV/c$^2$ in $\phi K^+K^-$, the branching fraction to pions divided
by that to kaons is inconsistent with the upper limit for the ratio
observed in Ref. \cite{okk} for $f_0(1710)$, this requires
two distinct resonances
$f_0(1790)$ and $f_0(1710)$.

\vspace{0.5cm} 
The BES collaboration thanks the staff of BEPC for
their hard efforts.  This work is supported in part by the National
Natural Science Foundation of China under contracts Nos.
19991480, 10225524, 10225525, the Chinese Academy of Sciences under
contract No. KJ 95T-03, the 100 Talents Program of CAS under Contract
Nos. U-11, U-24, U-25, and the Knowledge Innovation Project of CAS
under Contract Nos. U-602, U-34 (IHEP); by the National Natural Science
Foundation of China under Contract No.10175060 (USTC), No.10225522
(Tsinghua University); and the Department of Energy under Contract
No.DE-FG03-94ER40833 (U Hawaii).
We wish to acknowledge financial support from the Royal Society
for collaboration between the BES group and Queen Mary,
London under contract Q771.

\begin{figure}[htbp]
\centerline{\epsfig{file=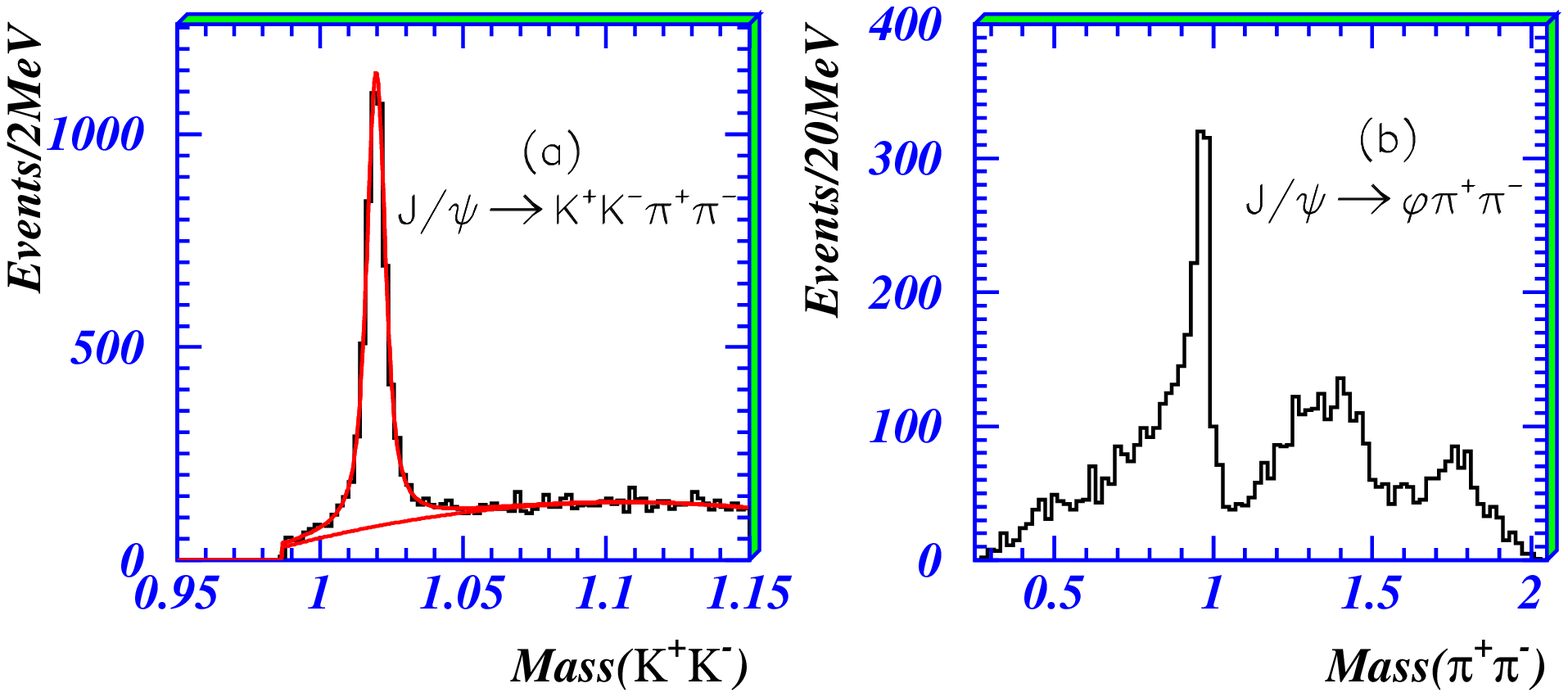,width=5.0in}}
\centerline{\epsfig{file=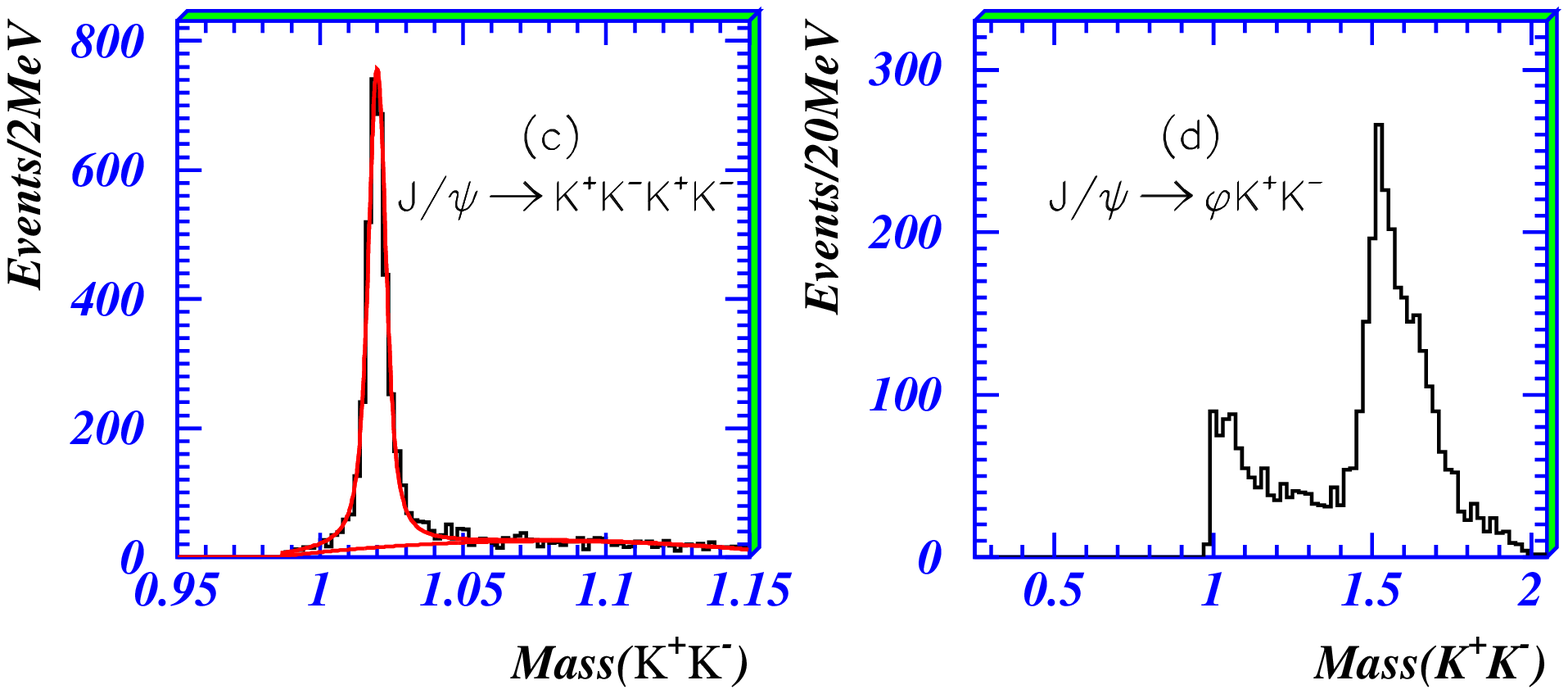,width=5.0in}}
\centerline{\epsfig{file=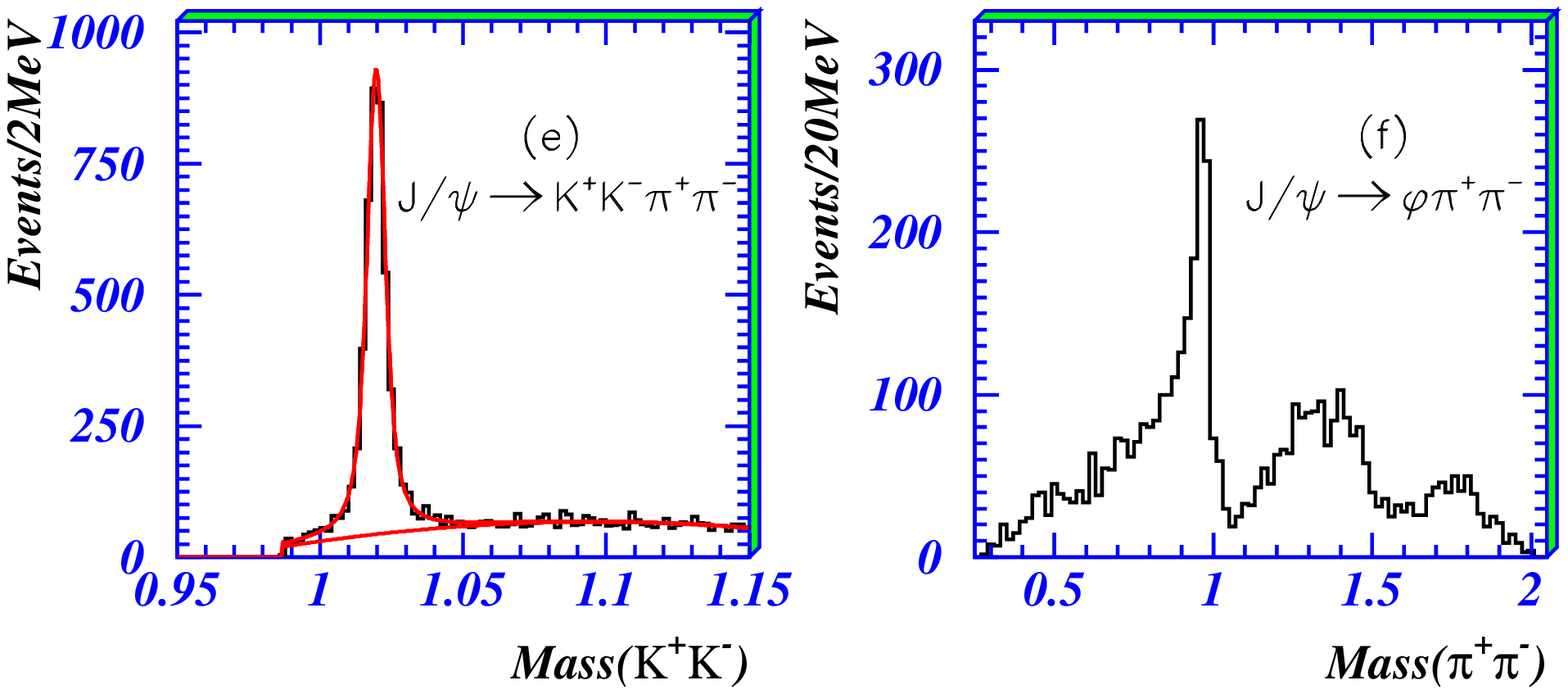,width=5.0in}}
\caption[]{The $K^+K^-$ invariant mass distributions for
(a) $J/\psi \to K^+K^-\pi ^+\pi ^-$,
(c) $J/\psi \to K^+K^-K^+K^-$;
curves show the fitted background and a Gaussian fit to the $\phi$;
(b) and (d) show mass projections for events selected within $\pm 15$
MeV/c$^2$ of the $\phi$;
(e) and (f) show mass projections after cutting
events within $\pm 100$ MeV/c$^2$ of the central mass of $K^*(890)$;
curves in (e) show the fitted background and a Gaussian fit to the
$\phi$. }
\label{figure1}
\end{figure}

\begin{figure}[htbp]
\centerline{\epsfig{file=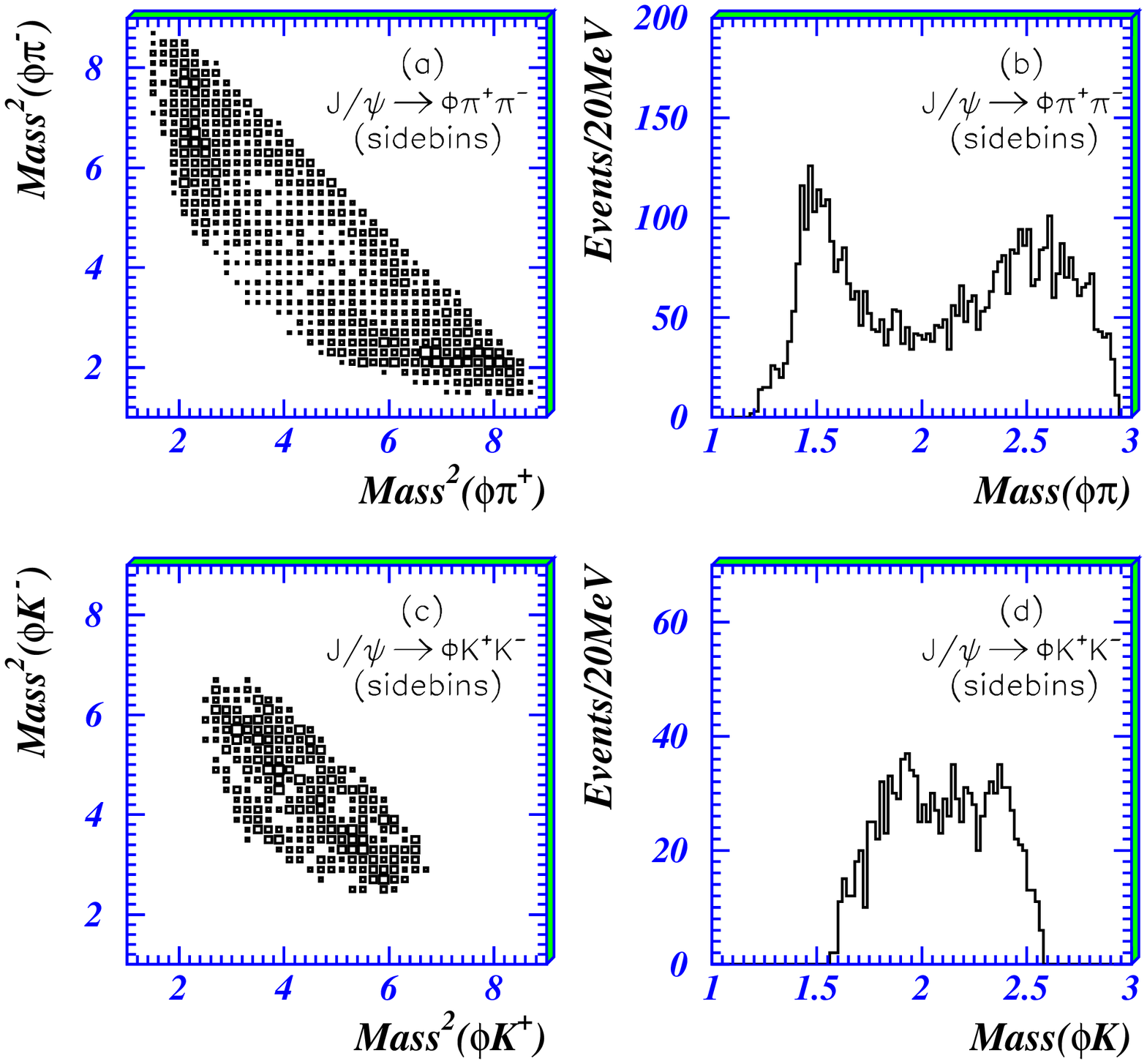,width=5.0in}}
\centerline{\epsfig{file=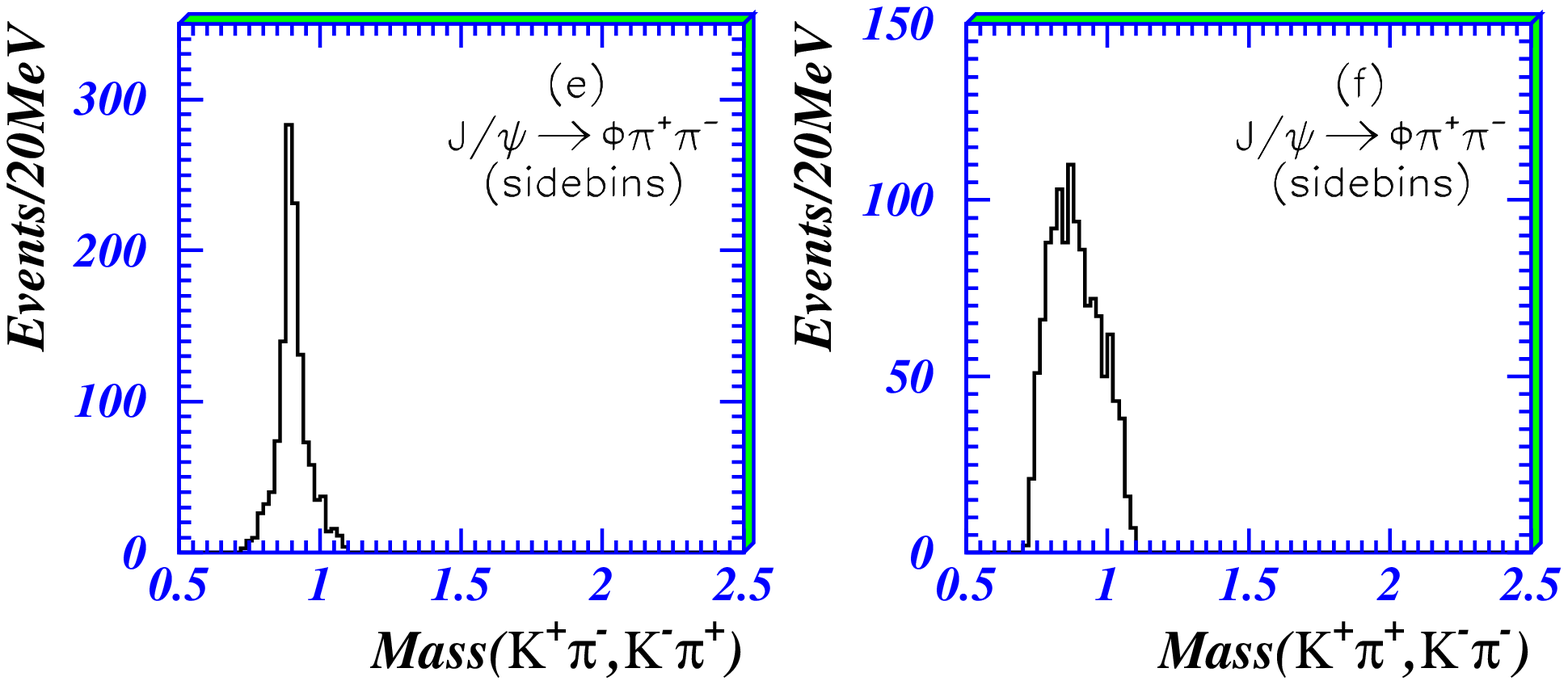,width=5.0in}}
\caption[]{Dalitz plots for a sidebin to the $\phi$ with $M(K^+K^-) = 1.045$
to 1.09 GeV/c$^2$ for (a) $K^+K^-\pi ^+\pi ^-$ data, (c) $K^+K^-K^+K^-$ data;
(b) and (d) show projections against $\phi \pi ^{\pm }$ and
$\phi K^{\pm}$ mass;
the mass distributions for (e)
$M(K^+\pi ^-)$ and $M(K^-\pi ^+)$,
(f) $M(K^+\pi ^+)$ and $M(K^-\pi ^-)$ for events from (a) in the
$\phi \pi$ mass range 1400--1600 MeV/c$^2$.}
\label{figure2}
\end{figure}

\begin{figure}[htbp]
\centerline{\epsfig{file=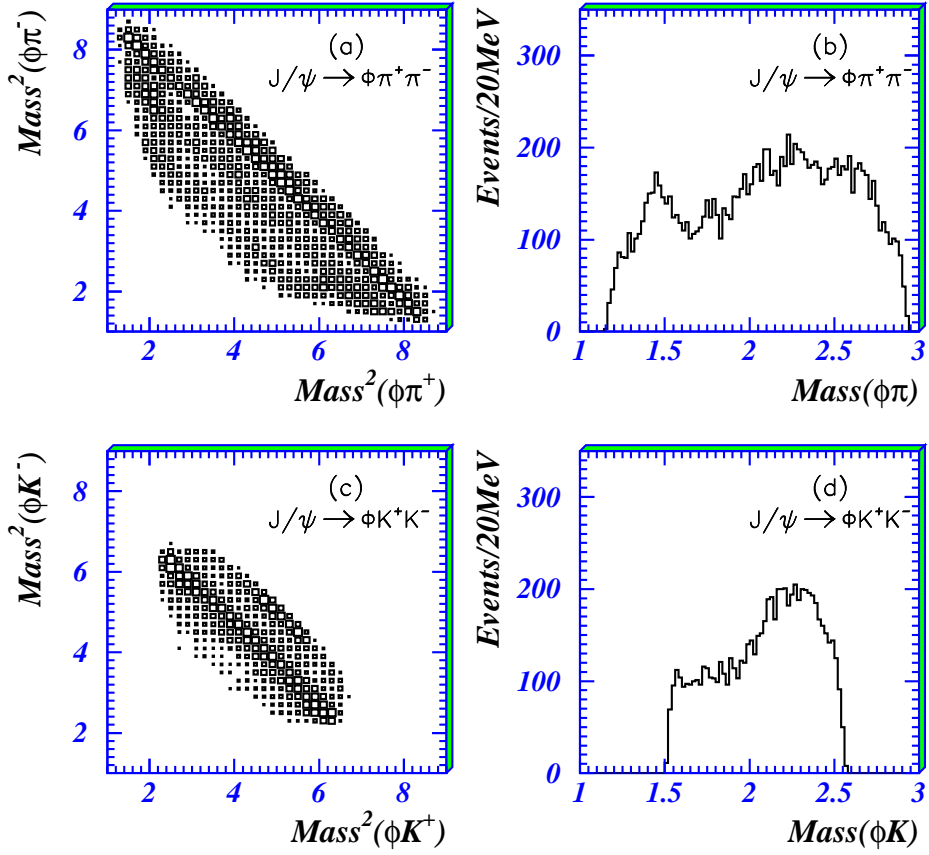,width=5.0in}}
\caption[]{Dalitz plots for
(a) $J/\psi \to \phi \pi ^+\pi ^-$,
(c) $J/\psi \to \phi K^+K^-$; (b) and (d) show the projections against
$\phi \pi$ and $\phi K$ mass.}
\label{figure3}
\end{figure}

\newpage
\begin{figure}[htbp]
\centerline{\epsfig{file=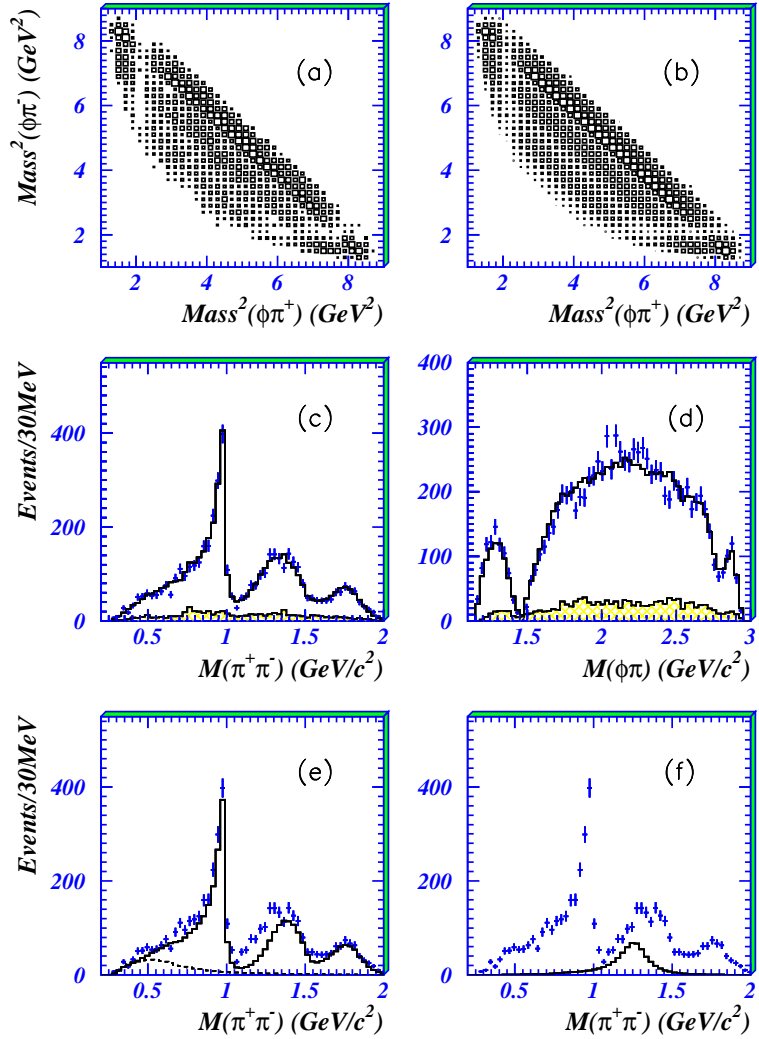,width=4.0in}}
\caption[]{(a) and (b) show measured and fitted Dalitz plots for
$J/\psi \to \phi \pi ^+\pi ^-$ after cutting $K^*(890)$ events.
(c) and (d) show mass projections;
the upper histogram shows the maximum likelihood fit and the lower
one shows background;
(e) shows the $f_0$ contribution from the fit (full histogram) and
the lower curve the $\sigma$ contribution, (f) the $f_2(1270)$
contribution. }
\label{figure4}
\end{figure}

\newpage
\begin{figure}[htbp]
\centerline{\epsfig{file=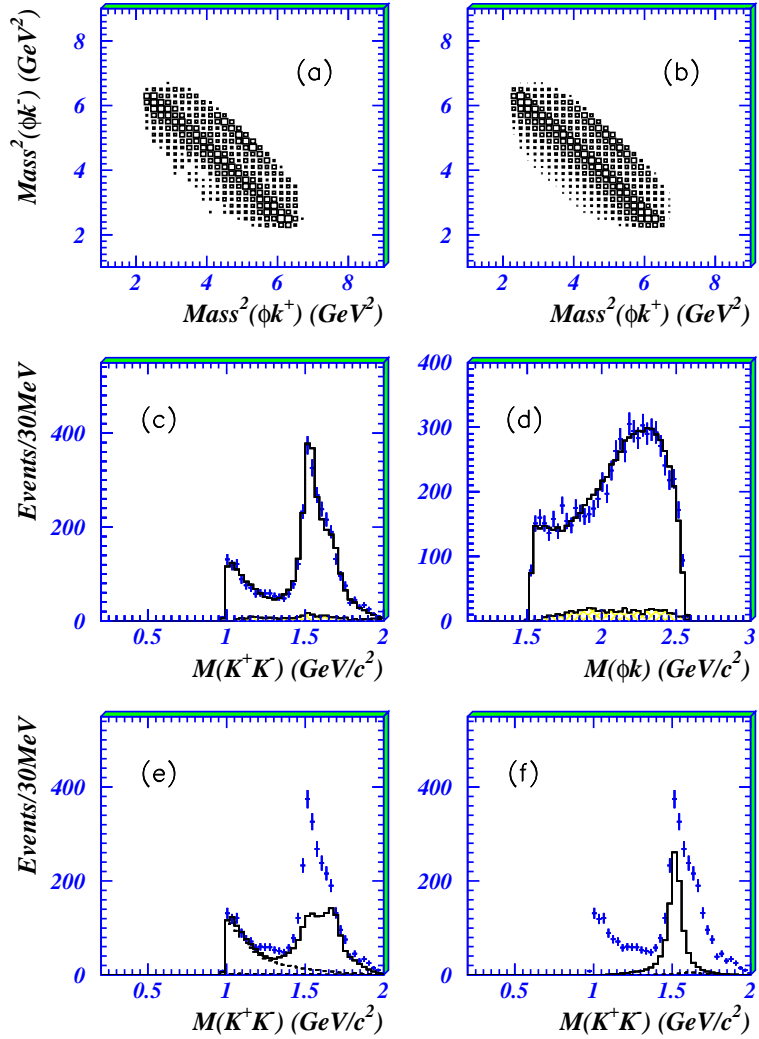,width=4.0in}}
\caption[]{(a) and (b) show measured and fitted Dalitz plots for
$J/\psi \to \phi K ^+K ^-$.
(c) and (d) Mass projections for $J/\psi \to \phi K^+K^-$ data,
compared with histograms from the fit;
(e) shows the $f_0$ contribution from the fit (full histogram)
and the lower curve the $f_0(980)$ contribution.
(f) the $f_2^\prime(1525)$ contribution from the fit (full histogram).}
\label{figure5}
\end{figure}

\begin{figure}[htbp]
\centerline{\epsfig{file=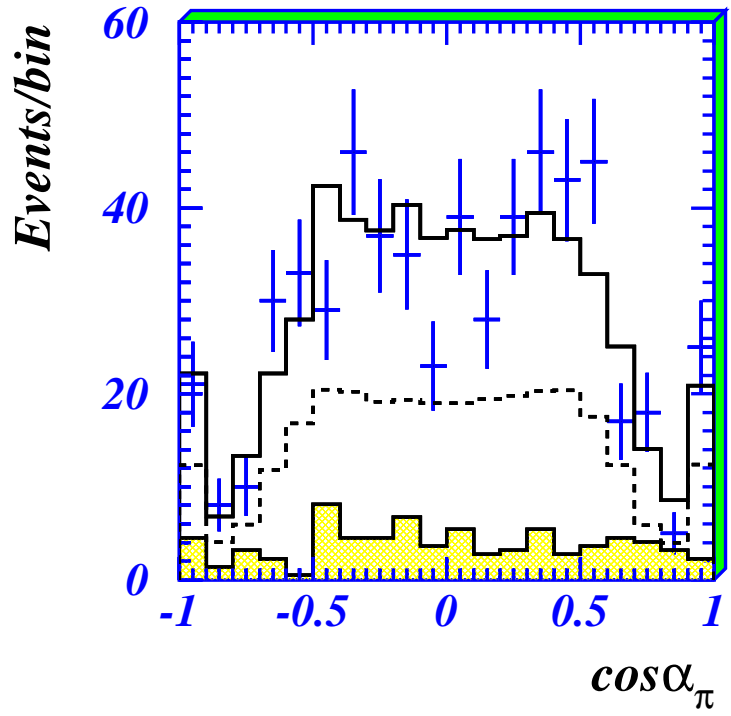,width=4.0in}}
\caption []{Angular distributions in $\phi \pi ^+\pi ^-$ data
(crosses) for $\alpha _\pi$, the angle of the $\pi ^+$ from $f_J$
decay with respect to the direction of $f_J$ in its rest frame.
The upper histograms shows the fit, and the lower one the
background. The dashed histogram shows the acceptance.}
\label{figure6}
\end{figure}

\end{document}